\pdfoutput=1
\documentclass[a4paper,fleqn]{cas-dc}

\usepackage[numbers]{natbib}

\def\tsc#1{\csdef{#1}{\textsc{\lowercase{#1}}\xspace}}
\tsc{WGM}
\tsc{QE}
\tsc{EP}
\tsc{PMS}
\tsc{BEC}
\tsc{DE}
\usepackage{graphicx}
\usepackage{epstopdf}
\begin{document}
\let\WriteBookmarks\relax
\def\floatpagepagefraction{1}
\def\textpagefraction{.001}
\shorttitle{Flexible magnetic filaments in rotating field}
\shortauthors{A.Zaben et~al.}

\title [mode = title]{Deformation of flexible ferromagnetic filaments under a rotating magnetic field}                      
%\tnotemark[1,2]

%\tnotetext[1]{This document is the results of the research  project funded by the National Science Foundation.}

%\tnotetext[2]{The second title footnote which is a longer text matter   to fill through the whole text width and overflow into   another line in the footnotes area of the first page.}

\author[1]{Abdelqader Zaben}

%\ead[url]{mmml.lu.lv}

%\credit{Conceptualization of this study, Methodology, Software}

\address[1]{MMML lab, University of Latvia, Jelgavas 3, Riga, LV-1004, Latvia}

\author[1]{Guntars Kitenbergs}

\author[1]{Andrejs C\={e}bers}[type=editor]
\cormark[1]
%\fnmark[1]
\ead{andrejs.cebers@lu.lv}

%\credit{Data curation, Writing - Original draft preparation}

\cortext[cor1]{Corresponding author}
%\cortext[cor2]{Principal corresponding author}
%\fntext[fn1]{This is the first author footnote. but is common to third author as well.}
%\fntext[fn2]{Another author footnote, this is a very long footnote and  it should be a really long footnote. But this footnote is not yet  sufficiently long enough to make two lines of footnote text.}

%\nonumnote{This note has no numbers. In this work we demonstrate $a_b$  the formation Y\_1 of a new type of polariton on the interface  between a cuprous oxide slab and a polystyrene micro-sphere placed  on the slab.  }

\begin{abstract}
Research on magnetic particles dispersed in a fluid medium, actuated by a rotating magnetic field, is becoming increasingly active for both lab-on-chip and bio-sensing applications. In this study, we experimentally investigate the behaviour of ferromagnetic filaments in a rotating field.Filaments are synthesized by linking micron-sized ferromagnetic particles with DNA strands. The experiments were conducted under different magnetic field strengths, frequencies and filament sizes, and deformation of the filaments was registered via microscope and camera. The results obtained showed that the body deformation is larger for longer filaments and higher frequencies. The angle between the filament tangent at the centre and the magnetic field direction increases linearly with frequency at the low-frequency regime. A further increase in the frequency will result in filament movement out of plane when the angle approaches 90 degrees. The experimental results were used to estimate the magnetic moment and the bending elasticity of the filament.  
\end{abstract}

%\begin{graphicalabstract}
%\includegraphics{figs/grabs.pdf}
%\end{graphicalabstract}

%\begin{highlights}
%\item Research highlights item 1
%\item Research highlights item 2
%\item Research highlights item 3
%\end{highlights}


\maketitle

\section{Introduction}
Flexible magnetic filaments are interesting for different applications, microfluidic mixing \cite{micromixing} and transport \cite{transport}, sensors \cite{1review}, creating self-propelling magnetic microdevices and others \cite{FilamentsReview}.
For these applications ferromagnetic filaments are of particular interest, which may be created by linking ferromagnetic microparticles \cite{JMMM2009}.
Several interesting phenomena are known for ferromagnetic filaments - in the AC magnetic field with a sufficiently high frequency they orient perpendicularly to the field \cite{JMMM2009}. 
When the magnetic field direction is changed, filaments form snake like 'S' shape \cite{Snake}.
If the direction is inverted, a ferromagnetic filament makes a loop \cite{ErglisMHD}, which further relaxes by 3D motion.
Due to the loop formation it is possible to create self-propelling magnetic microdevices \cite{PRE2009}. Considerable effort has been put to numerical modelling of flexible magnetic filaments. For example, the behavior of small filaments under constant field has been investigated in the dilute regime \cite{Kantorovich1} and under external flow \cite{Kantorovich2}. Dynamics of a filament in rotating field have been studied in \cite{PRE2017}.
In the present work we investigate the physical properties of ferromagnetic filaments under a rotating magnetic field, focusing on the deformation dependence on the frequency and strength of the field.

\section{Experimental}

\begin{figure*}
    \centering
    \includegraphics[width=1\textwidth]{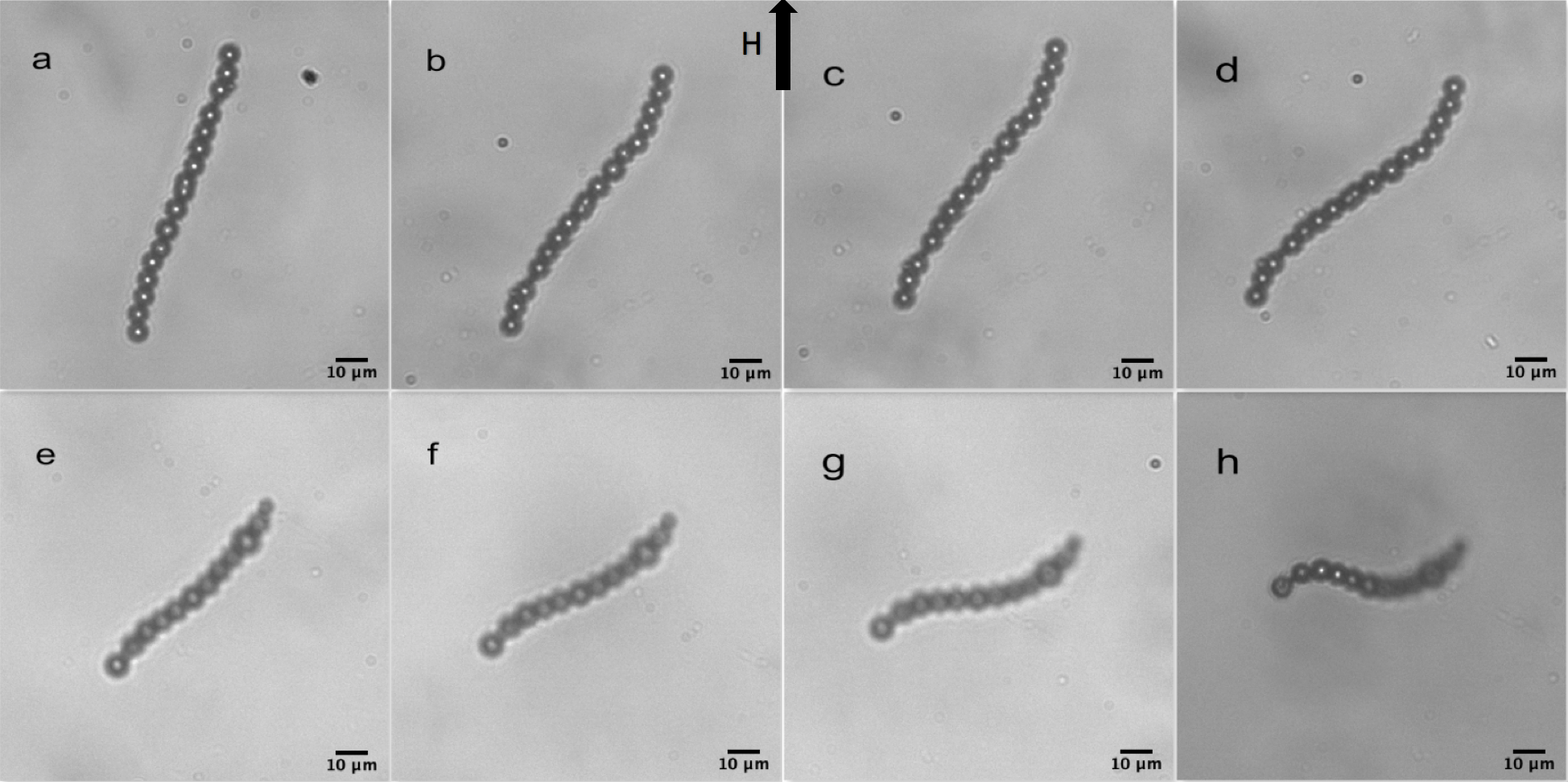}
    \caption{Behavior of two flexible magnetic filaments under
rotating magnetic field $H=8.6$~Oe. Filament with $L=67.4~\mu$m at (a) $0.2$~Hz, (b) $0.3$~Hz, (c) $0.4$~Hz and (d) $0.6$~Hz. Filament with $L = 50.5~\mu$m at (e) $1.0$~Hz, (f) $1.5$~Hz, (g) $2.0$~Hz and (h) $3.0$~Hz. In (h), the filament moves out of imaging plane.}
    \label{fig:1}
\end{figure*}

\subsection{Synthesis}
The filament synthesis methodology was adapted from   \cite{K.thesis}.
In this way filaments with a length $L$ between $8~\mu$m to $80~\mu$m are formed. 
For example, see Fig.\ref{fig:1}(a). 
The filaments were prepared by mixing $4.26$~$\mu$m large ferromagnetic particles (Spherotech, $1\%$w/v) with DNA strands (ASLA biotech, $182~\mu$g/ml) in TE buffer ($10\%$)  solution.
Ferromagnetic particles are made from polymer beads that are covered with chromium dioxide.
Their surface is functionalized with streptavidin.
DNA strands are 1000bp long and have biotin at their ends.

Each sample was made by mixing $0.5$~ml TE buffer solution ($\eta = 0.01$~P, pH = 7.5) with 10 $\mu$L of DNA solution and $2~\mu$L particle suspension.  This corresponds to a resulting DNA concentration of $6~nM$ and particle mass concentration of $0.004\%~w/v$.
The sample is then quickly placed in the middle of two $5\times5\times1$~cm Neodymium magnets, fixed $7$~cm apart, providing a homogeneous field of $\approx500$~Oe.
It is left in the field for two minutes, allowing the particles to form chains.
While chains are being formed, the biotin ends of the DNA find their way to the streptavidin surface of beads and link the particle chains permanently.On average, each particle has $\approx8\cdot10^6$~linkers, which is slightly less than their maximum capacity in \cite{K.thesis}.

\subsection{Video microscopy}

Filaments were observed with an optical microscope (Leica DMI3000B) using a $40\times$ objective in bright field mode.
The magnetic field is generated using a coil system that consists of six coils placed to provide fields in three directions. An AC power supply (Kepco BOP 20-10M)  connects with the coil pairs in each direction, which is capable of producing fields of up to $120$~Oe, with a maximum frequency of $50$~Hz.
The current output is controlled using National Instruments data acquisition card and LabVIEW code.
It generates analogue sine and cosine signals for a rotating magnetic field.
The fluidic cells are prepared using two glass slides separated by $130$~$\mu$m double-sided adhesive tape.
Each cell is formed by a $1 \times 1$~cm cut in the tape, in which $10~\mu$l of the prepared solution are added.
The images are taken using Basler ac1920-155um camera.
Depending on the measurement, it is used either in a fixed frame rate, for which a maximum of $200$~fps can be achieved, or by a trigger mode, which allows to synchronize it with the magnetic field direction. 

\subsection{Image processing}
\label{sec:2.3}
\begin{figure} 
    \centering
    \includegraphics[width=\columnwidth]{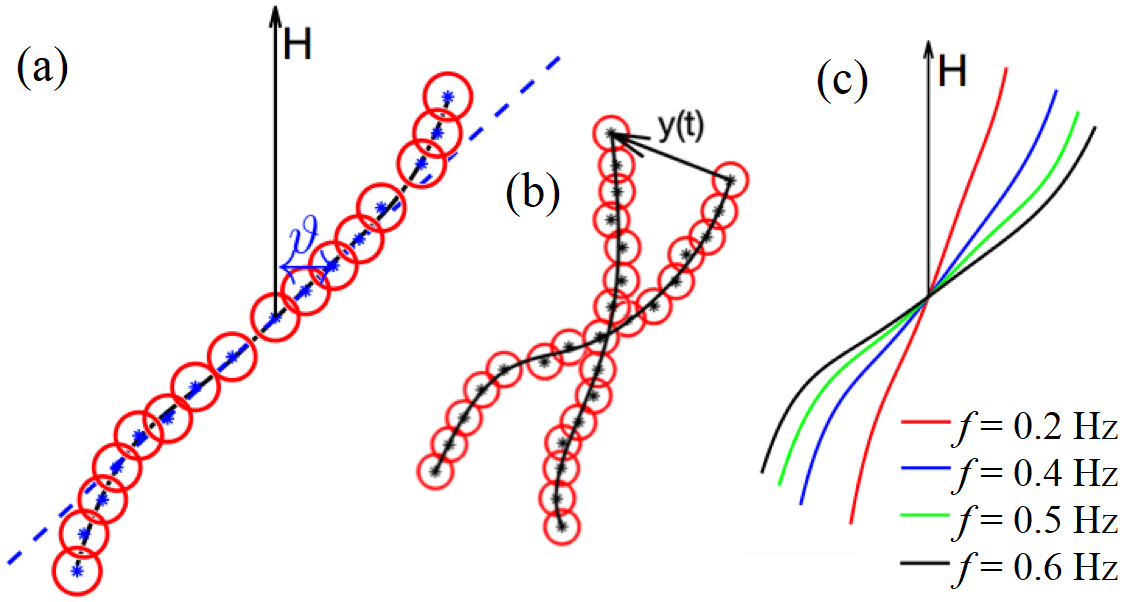}
    \caption{(a) An illustration of filament image processing. Polynomial fit (black curve) of centres of particles (blue asterisks and red circles) describes the deformed filament. The angle ($\vartheta$) between the tangent at the filament center (blue dashed line) and the magnetic field direction (indicated with the black arrow) is used for characterizing deformation. (b) An example of deformation relaxation measurement. The position of filament tip particle (y-discplacement) is tracked over time. (c) Experimental magnetic filament deformations for different frequencies $f$ ($0.2$, $0.4$, $0.5$ and $0.6$~Hz) at $H=8.6$~Oe. }
    \label{fig:2}
\end{figure}

The images used for characterising the filament deformation were obtained by trigger mode in such a way that the magnetic field points to the north, as shown in Fig.\ref{fig:1}. Deformation of the filament may be characterized by the measurement of the tangent angle $\vartheta$ at the centre (shown in Fig.\ref{fig:2}(a)) and at the tip, which are compared with the results of the numerical simulations. The tangent angle at the tip was found to be not along the field; however close to it (data not shown). 

Image processing was done with Matlab and the steps are depicted in Fig.\ref{fig:2}(a).
Initially, the particle centres (blue asterisks) are detected using a circular object detection function (red circles).
Then the shape of the filament is approximated by fitting the particle centre coordinated with a polynomial of sixth order (black curve).
This allows us to find the tangent at the centre (blue dashed line) and calculate the angle $\vartheta$. 

Experiments for analysing the deformation relaxation time were obtained by setting the frame rate to $150$~fps. Initial deformation of the filament is made by a rotating magnetic field having strength values between $8.6$~Oe and $17.2$~Oe. 
Then the magnetic field is turned off and the relaxation is observed, while tracking the displacement of the filament tip particle.
An example of processed image for tracking the displacement of tip particle $y(t)$ is shown in Fig.\ref{fig:2}(b).
\section{Results and Discussion}
The experiments were conducted in the following manner.
The fluidic cell is placed for observation under the microscope.
Due to gravity effects, the filaments sediment to the bottom of the fluidic cell.
A static magnetic field of $17.2$~Oe is applied in the vertical direction for the initial alignment of the filament. A further static field is switched off, and a rotating field is applied in the clockwise direction. the field strength is between $6$~Oe and $25$~Oe, and a frequency up to $8$~Hz.
The image acquisition is set to trigger mode.
Examples of steady-state configurations of two filaments in the rotating field are shown in Fig.\ref{fig:1}. Characteristic S-like shape may be seen, which is also observed for chains of paramagnetic particles, see for example \cite{PhysRevE.69.041406}.

Increase of frequency induces a larger deformation of the filament.
This can be clearly seen in Fig.\ref{fig:2}(c), where the filament polynomial fits for various frequencies ($0.2$~Hz (red curve), $0.4$~Hz (blue curve), $0.5$~Hz (green curve) and $0.6$~Hz (black curve)) at fixed magnetic field strength $H=8.6$~Oe are compared.

\begin{figure}
    \centering
    \includegraphics[width=\columnwidth]{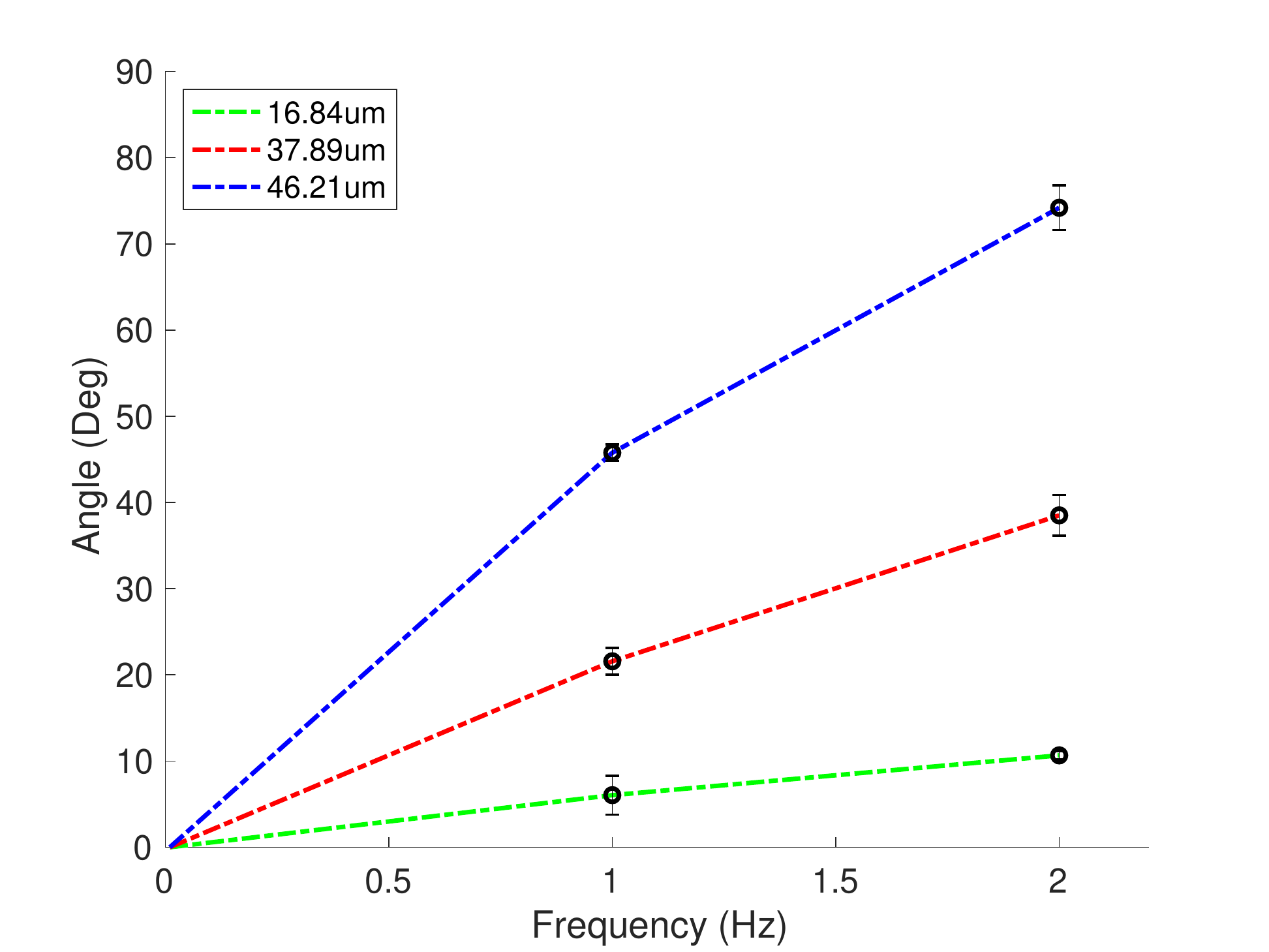}
    \caption{Relationship between the angle $\vartheta$ and frequency under rotating magnetic field, $H$ = $6.9$~Oe, for filaments with different lengths: $L$ = $46.2$~$\mu$m (blue line), $L$ = $37.9$~$\mu$m (red line) and $L$ = $16.8$~$\mu$m (green line).}
    \label{fig:3}
\end{figure}

Measurements were done for filaments with different lengths. 
The results show that longer filaments have higher angle $\vartheta$ at a fixed frequency and field strength.
Hence, longer filaments tend to deform more, while having more flexibility to deform by viscous torques under a rotating field.
Angle as a function of the frequency for different filament lengths is shown in Fig.\ref{fig:3}.

\begin{figure}
    \centering
    \includegraphics[width=\columnwidth]{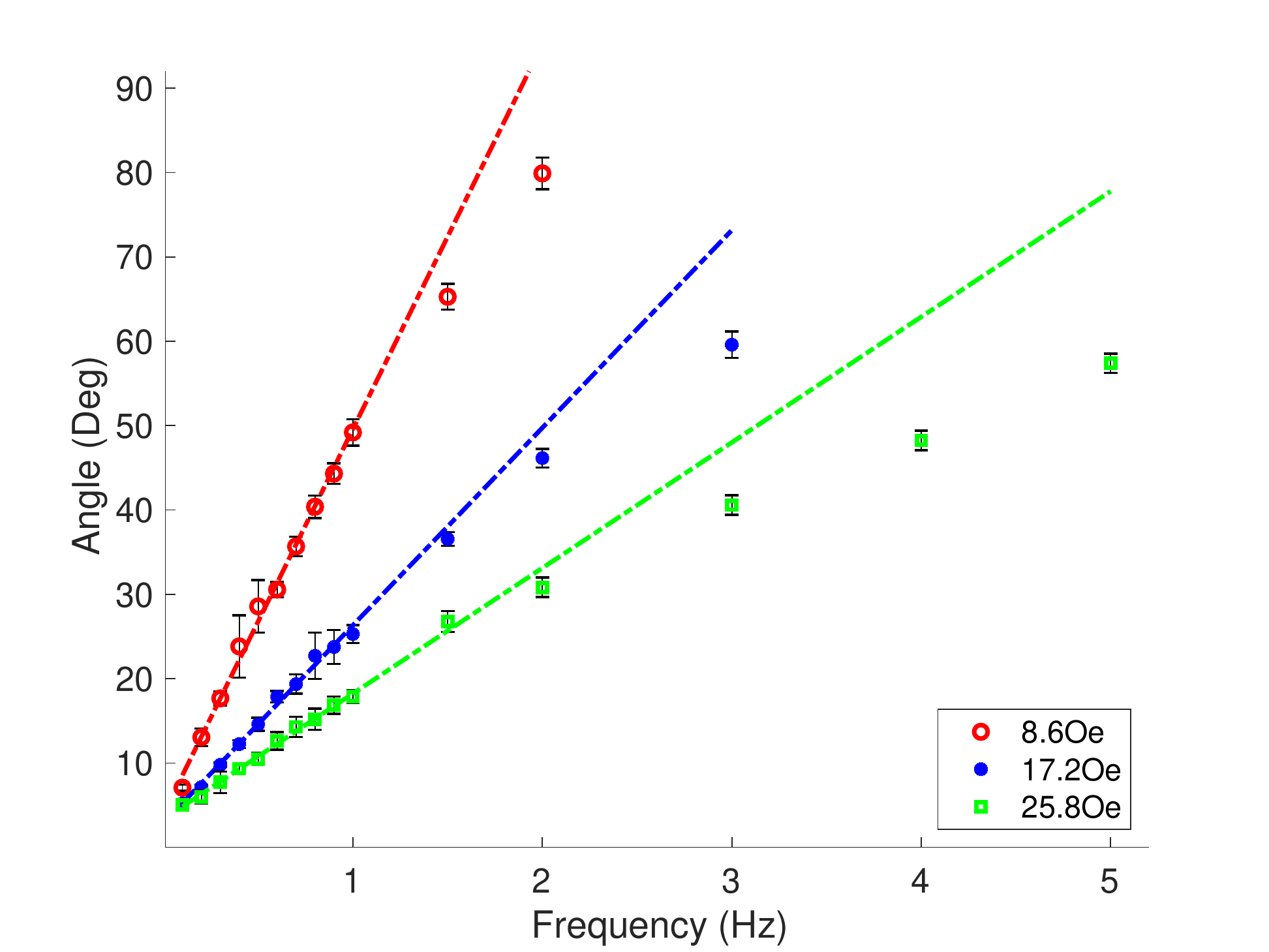}
    \caption{Relationship between the angle $\vartheta$ and frequency, under rotating magnetic field with different strengths, $H$ = $8.6$~Oe (red line), $H$ = $17.2$~Oe (blue line) and $H$ = $25.8$~Oe (green line), filament length $L$ = $46.3$~$\mu$m.}
    \label{fig:4}
\end{figure}

The results for the effect of varying magnetic field strength $H$ and frequency $f$ was obtained using a single filament for each analysis.
The rotating magnetic field introduces a magnetic torque, which is a function of the field strength opposed by viscous torque, dependent on the length $L$ and the angular velocity $2\pi f$.
The increase of the frequency at a constant $H$ increases the characteristic angle $\vartheta$.
Results are shown in  Fig.\ref{fig:4}.
At low frequencies this relationship is found to be linear.
A further increase in the frequency tends to lower the gradient as a result of the increased viscous forces.
The filament tends to deform having 'S' like shape as visible in Fig.\ref{fig:1}(g), where the filament centre tends to orient perpendicular to the magnetic field direction.  

The effect of increasing the field strength $H$ at constant frequency decreases the deformation and angle $\vartheta$, due to the increased torque to overcome the friction forces.

In these experiments filaments were found to move synchronously with the magnetic field (data not shown).
This is different than observed in numerical modelling by \citet{PRE2017}.
There, if frequency is increased, the filaments reach an asynchronous regime, where the filaments experience back and forth motion due to lag in following the magnetic field.
By comparison, in experiments a further increase in $\vartheta$ results in filament movement out of the image plane, in the third dimension.
This transition was noted to occur when the filament centre orients perpendicularly with the field direction, having  $\vartheta$ of  90$^{\circ}$.
An example of this behavior visible in figure Fig.\ref{fig:1}(h).
As a result, we can assume that small fluctuations deform the filament to a three dimensional shape, which is dynamically preferable, before it reaches an asynchronous regime.

Similar deformed 'S' like filament shapes with back and forth motion can be observed for paramagnetic filaments in specific rotating magnetic field conditions \cite{ParamagneticFilaments}.
However, the difference in magnetic properties does not allow to compare the behavior quantitatively.

For our experiments we can describe the angle $\vartheta$ dependence by a theoretical model for filament dynamics under rotating field presented in \citet{ErglisMHD}, where numerical simulations give the following relation
 \begin{equation} \label{eq:1}
\vartheta = 0.086 \frac{\omega\tau}{Cm}, \ 
\end{equation}
where $\tau$ is the elastic deformation time $ \tau =  \frac{\zeta L^4}{A_{b}}$ , $\omega$ is the angular velocity $\omega=2\pi f$ and $Cm$ is the magnetoelastic number, $Cm =  \frac{MHL^2}{A_{b}} $ for which $L$ is the filament length, $H$ is the magnetic field strength, $A_{b}$ is the bending modulus and $\zeta$ is the hydrodynamic drag coefficient which is estimated by 4$\pi\eta$.

\begin{figure}
    \centering
    \includegraphics[width=\columnwidth]{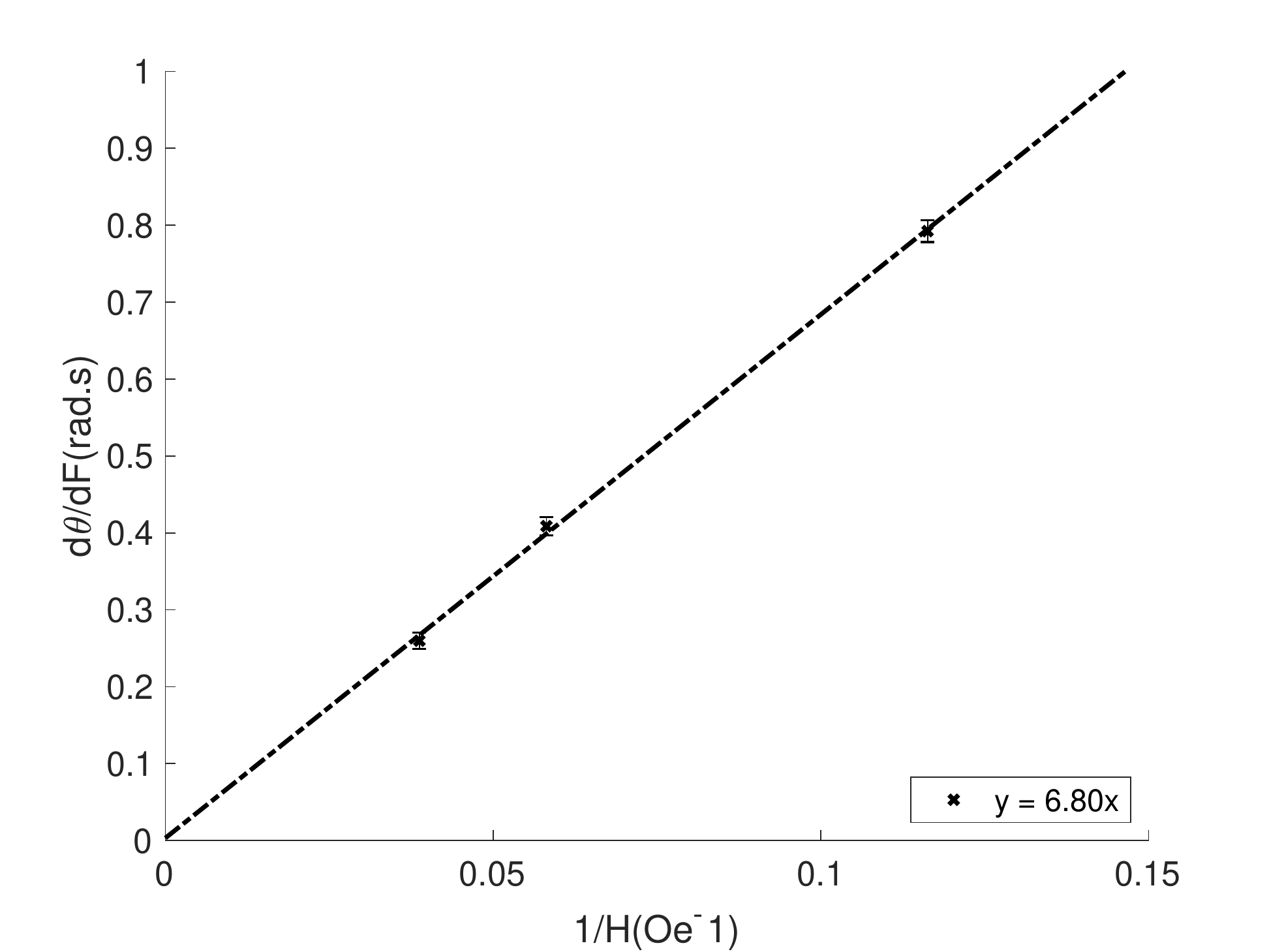}
    \caption{Relationship between  \( \frac{d\vartheta}{df} \) and  \( \frac{1}{H} \), obtained for a filament with length $L =46.3$~$\mu$m.}
    \label{fig:5}
\end{figure}

We use relation \eqref{eq:1} to analyze the linear regimes in Fig.\ref{fig:4}.
Slopes $\frac{d\vartheta}{df}$ as a function of $\frac{1}{H}$ for a filament $L=46.3~\mu$m are shown in Fig.\ref{fig:5}.
The experimental data is fitted  with a linear relation $\frac{\vartheta}{f} = a\,\frac{1}{H}$\,, for which we find $a = 6.8$~Oe$\cdot$s.
From this we find information about filament magnetization.
For magnetization per unit length we get $M = 2.14\cdot10^{-7}$emu and for the magnetic moment of particle $m$ = $9.01\cdot10^{-11}$emu $($M$ \cdot d)$. 
The last value is not far from estimates in \cite{ErglisMHD}.

\begin{figure}
    \centering
    \includegraphics[width=\columnwidth]{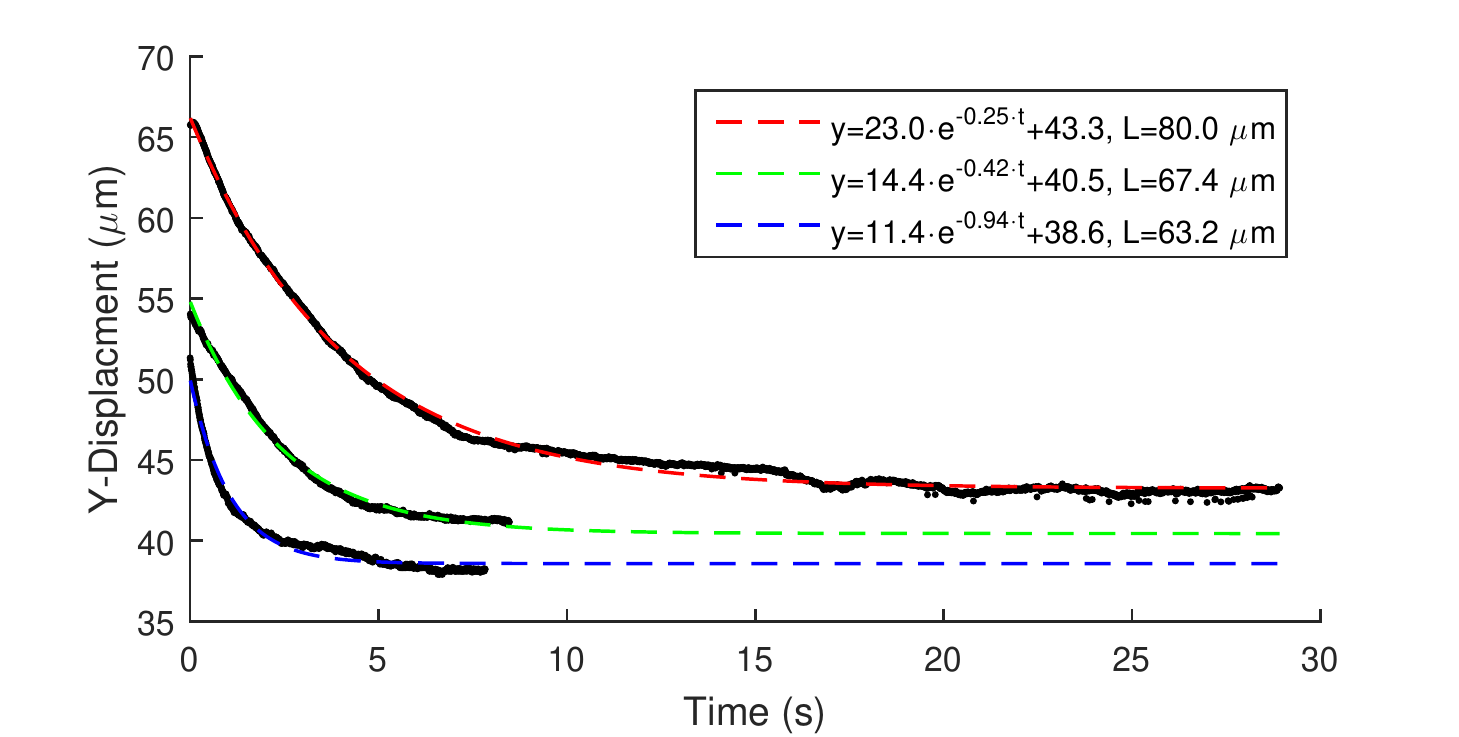}
    \caption{Filament relaxation behaviour for different filament length and initial rotating field conditions: $L = 80.0$~$\mu$m, $f = 1.0$~Hz and $H = 8.6$~Oe (red curve). $L = 63.2$~$\mu$m, $f = 1.0$~Hz and $H = 13.7$~Oe (blue curve). $L = 67.4$~$\mu$m, $f = 1.5$~Hz and $H = 8.6$~Oe (green curve).}
    \label{fig:6}
\end{figure}

In order to measure the bending modulus, the protocol for filament relaxation was carried out, as explained in \S\ref{sec:2.3}.
The motion of the filament tip shows an exponential time dependence, as can be seen in Fig.\ref{fig:6}.
The decrements of relaxation are different for different filament length and are found to be equal to $0.25$~s$^{-1}$ for $L=80.0~\mu$m, $0.42$~s$^{-1}$ for $L=67.4~\mu$m and $0.94$~s$^{-1}$ for $L=63.2~\mu$m.
It is worth to note that in our experiments the filament does not return to its original shape, as visible in Fig.\ref{fig:2}(b).
Small remaining deformation may be the result of damaged bonds between the particles or due to wall surface drag.  

\begin{figure}
    \centering
    \includegraphics[width=\columnwidth]{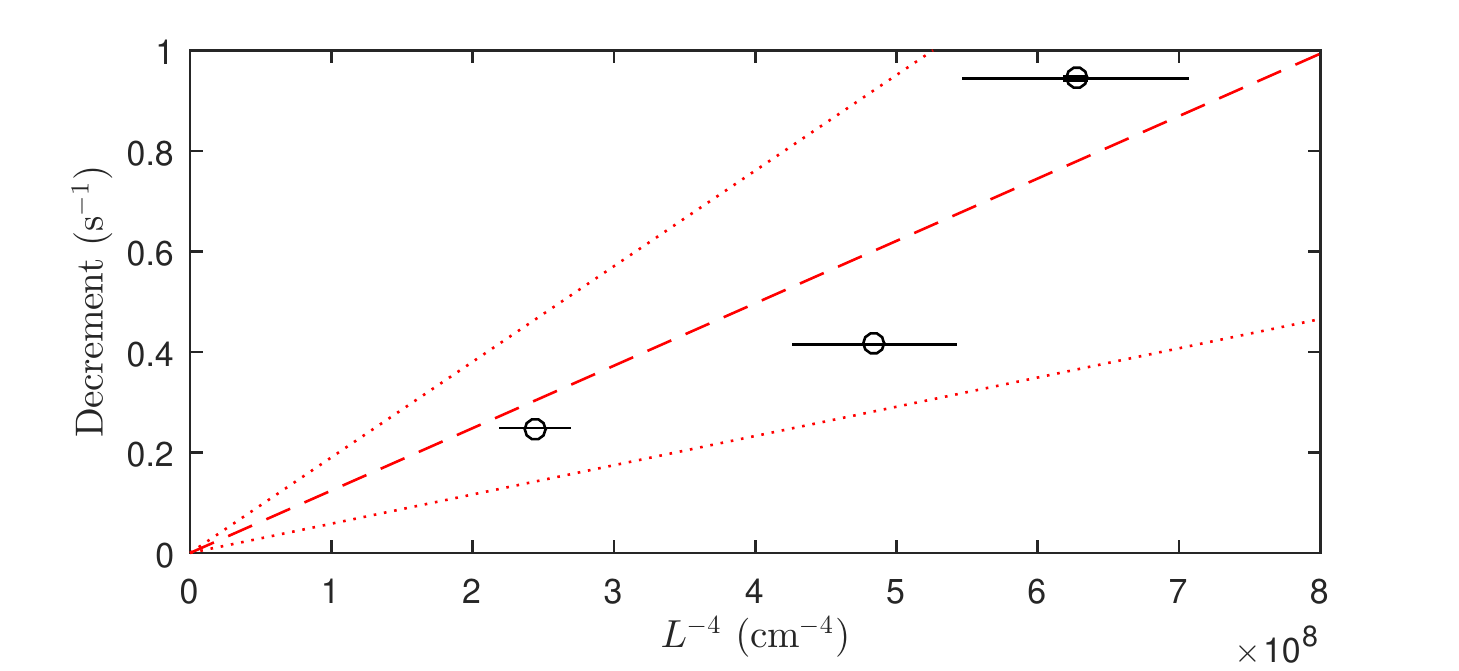}
    \caption{Relationship between relaxation decrements and $L^{-4}$ for different filament lengths. Black circles are experimental points with errorbars. Red dashed line is linear fit and red dotted lines are confidence intervals for $3\sigma$.}
    \label{fig:7}
\end{figure}

The filament relaxes due to the action of bending elasticity and is slowed down by viscous forces.
The solution of the elasticity problem for the rod at free and unclamped boundary conditions \cite{Goldstein} gives the spectrum of relaxation rates from which the smallest decrement is given by relation $3.93^{4}L^{-4}A_{b}$/$\zeta$. 
Taking experimental values of decrements and fitting as a function of $L^{-4}$ allows to estimate the average bending modulus of the rod.
This is done in Fig.\ref{fig:7}.
The fitted line and its error gives $(A_{b}\pm\Delta A_{b}) = (6.5\pm3.4)\cdot10^{-13}$ erg$\cdot$cm.
This is close to the estimate of the bending modulus obtained in \cite{JMMM2009}. The value obtained by relaxation experiments is by almost two orders of magnitude larger than the calculated by the relation for the dipolar interaction contribution to the bending modulus $A_{b} = M^{2}/2 $ given in \cite{FilamentsReview,Kiani_2015} (For more exact nevertheless close coefficient before $M^{2}$, see \cite{vella_pontavice_hall_goriely_2014}). Thus we may conclude that bending elasticity is determined by the linkers. In \cite{doi:10.1021/la5009939}, the bending modulus of chains of paramagnetic particles was measured by the study of thermal fluctuations of their shape. obtained results show that these chains, possibly due to the much smaller size of the particles, are significantly more flexible.

\section{Conclusions}

Properties of ferromagnetic filaments synthesized by linking funtionalized ferromagnetic
microparticles by biotinized DNA fragments are measured by the study of their deformation in a rotating magnetic field.
It is found that filaments at low frequencies have 'S' like
shapes and the tangent angle at the center of filament is proportional to the frequency and
inverse value of the field strength.
The fit of experimentally obtained dependence by the relation obtained numerically gives the magnetic moment of the ferromagnetic particles.
Very interesting observation was made at higher frequencies of the rotating field showing
that the filament is going out of the plane of rotating field.
Detailed study of this important phenomenon is pending for future publications.
The bending modulus of the filament is measured by registering its relaxation to straight configuration from initially prepared deformed shape.
The results obtained show that synthesized ferromagnetic filaments are quite stiff and their persistence length has the order of magnitude of several tenths of centimeters.

\section*{Acknowledgements}
A.Z. acknowledges support from the European Union\textquotesingle s Horizon 2020 research and innovation programme under grant agreement MAMI No. 766007, A.C. from M.era-net project FMF No.1.1.1.5./ERANET/18/04 and G.K. from PostDocLatvia grant No. 1.1.1.2/VIAA/1/16/197.

\printcredits

%% Loading bibliography style file
%\bibliographystyle{model1-num-names}
\bibliographystyle{model1-num-names}

% Loading bibliography database
\bibliography{cas-refs}

%\vskip3pt
\end{document}